\documentclass[11pt,a4paper]{article}
\usepackage{graphicx}
\usepackage{amsmath}
\usepackage{pdfpages}

\usepackage[numbers,super,sort&compress]{natbib}

\usepackage{amssymb}
\usepackage{url}
\usepackage{authblk}
\usepackage{float}

\usepackage[top=1.1in,left=2.75in,footskip=1in,marginparwidth=1in]{geometry}

\usepackage[left]{lineno}

\usepackage{microtype}
\DisableLigatures[f]{encoding = *, family = * }

\usepackage{changepage}

\usepackage{lastpage,fancyhdr} 
\usepackage{epstopdf}
\fancyheadoffset[L]{3.25in}
\fancyfootoffset[L]{2.25in}

\usepackage{color,soul}

\definecolor{Gray}{gray}{.25}

\usepackage{graphicx}

\usepackage[fulladjust]{marginnote}

\newcommand{\tr}{}
\DeclareRobustCommand{\tra}[1]{{\sethlcolor{white}{\hl{#1}}}}
\usepackage[hidelinks]{hyperref}
\usepackage{xcolor}
\hypersetup{
    colorlinks,
    linkcolor={red!50!black},
    citecolor={blue!50!black},
    urlcolor={blue!80!black}
}

\makeatletter
\newcommand{\settitle}{\@maketitle}
\makeatother

\date{}
\begin{document}

\title{\Large
\textbf
{{{Ordering and topological defects in social wasps' nests}}}}

\author[ ]{Shivani Krishna\textsuperscript{1,*}, 
Apoorva Gopinath\textsuperscript{1} and
Somendra M. Bhattacharjee\textsuperscript{2} }
\affil[1]{{\small{Department of Biology, Ashoka University, Sonepat 131029, India}}}
\affil[2]{{\small{Department of Physics, Ashoka University, Sonepat 131029, India}}}
\affil[ ]{\vspace{4pt}{\small*Corresponding author: shivani.krishna@ashoka.edu.in}}

\settitle
\thispagestyle{empty}
\addtocounter{page}{-1}

\pagebreak

\maketitle
\vspace{-0.2in}
\begin{abstract}
  Social insects have evolved a variety of architectural formations.
  Bees and wasps are well known for their ability to achieve compact
  structures by building hexagonal cells. {\it{Polistes wattii}}, an
  open nesting paper wasp species, builds planar hexagonal structures.
  \tr{Here, using the pair correlation function approach, we show that
  their nests exhibit short-range hexagonal order (no long-range
  order) akin to amorphous materials. Hexagonal orientational order was
  well preserved globally.} We also show the presence of topological defects such as dislocations (pentagon-heptagon
  disclination pairs) and Stone-Wales quadrupoles, and discuss how 
  these defects were organized in the nest, thereby restoring order. Furthermore, we suggest the possible role of such
  defects in shaping nesting architectures of other social insect
  species.

\end{abstract}
\section*{Introduction}
Animal nests and their architecture are crucial for the development of
the young, protection from predators, and storage of resources. The
diversity and complexity in architecture signify how these functions
are fulfilled in diverse habitats and environmental conditions. Such
complexity is exemplified by the nests of social insects (ants,
bees, termites, and wasps) owing to their ability to use or modify the
surroundings, and create ordered and plastic structures\cite{camazine_self-organization_2001,tschinkel_nest_2004,
  reid_army_2015}.  These nests have captivated mathematicians and
ecologists alike for their structural complexity and the mechanisms
behind the coordinated building\cite{grasse_termitology_1984,theraulaz_mechanisms_1999,
  hansell_animal_2005,peters_collective_2019}.  How do multiple
individuals in a group or a colony come together to construct these
tiled layouts? The principle of stigmergy explains this as the
formation of patterns by emergence
\cite{grasse_reconstruction_1959}, including self-organisation and
self-assembly mechanisms
\cite{theraulaz_coordination_1995,bonabeau_self-organization_1997}. In most species, the construction rules depend on environmental (e.g., humidity, soil moisture) or pheromone gradients\cite{camazine_self-organization_2001,khuong_stigmergic_2016}.  In
this context, the relative importance of self-organised\cite{penzes_round_1993, karsai_decentralized_1999} vs
template-based interactions has been examined, and current evidence
points towards architectural mechanisms being an interactive effect of
both these interactions\cite{perna_when_2017}.  It is paramount to
place these mechanisms in the context of behavioural and cognitive
processes to understand nest construction comprehensively. The
repertoires required for complex nest architectures are often
characterised by simplified clusters of direct steps. However, the
processes involved in planning construction in various environmental
conditions, detecting errors and their remediation, working around
novel obstructions, etc., require substantial cognitive abilities that
outwit simple algorithms\cite{gallo_cognitive_2018}.

In bees and wasps, cells within the nest are usually circular or
hexagonal. Hexagonal tiling provides a compact structure and is proven to cover a
planar region with regular units of equal area whilst minimising the
total perimeter (honeycomb conjecture\cite{hales_honeycomb_2001}).
Such optimal utilisation of space and energy has been proposed as the
significant selection pressure for shaping these structures
\cite{toth_what_1964,jeanne_adaptiveness_1975,karsai_optimality_1999,gallo_cognitive_2018}.
Honeybees use wax to shape their cells, while most wasp species use
either mud or fibrous materials of plant origin for construction.
Though the cells are hexagonal in both cases, the construction
materials possess strikingly distinct properties. Therefore, the
underlying processes could be different. Construction of bees'
wax-based cells is postulated to begin as an array of circles (laid
over packed cylinders as a base) that are modified to rounded hexagons
purely by mechanical/thermodynamic means\cite{pirk_honeybee_2004,karihaloo_honeybee_2013}. However,
experimental work from nests of European honeybees suggests that bees
actively construct hexagonal cells by handling the wax and controlling
its temperature\cite{bauer_hexagonal_2013}. Given the plasticity of
wax, any disruptions to tiling in these structures can be rectified,
\tr{thereby reducing the possibility of overall defects in the nests. Hence the legend of honeycomb ordering}.
On the other hand, nests made up of plant fibres such as those of
paper wasps are unlikely to follow similar modes of construction. The
reduced plasticity makes them interesting as ordered physical systems
prone to topological defects. A topological defect in systems of
broken symmetry, like crystals, magnets, liquid crystals, etc., is
defined as a disruption of order throughout the system in a way that
defies restoration via any continuous deformation\cite{mermin_topological_1979, bhattacharjee_use_2017}.  In contrast,
defects such as vacancies, hilly terrain in a flat land, etc., are
geometric defects as they affect order locally in their neighbourhood
without any signature far away from them. A key characteristic of
topological defects is that they cannot be repaired by local
rearrangements alone but instead require changes to the system
globally. Geometric defects can be amended by making changes
locally at the scale of basic elements or constituents\cite{griffin_relationship_2017}.
Here, we refer to { {topological defects}} as those that occur and affect at length scales
beyond a single unit/element, such as a single cell within a nest. The
emergence and evolution of such topological defects in paper wasps'
nests have never been studied. Some of the most common topological defects in crystals and liquid
crystals manifest as disclinations and dislocations\cite{harris_disclinations_1977,de_gennes_physics_1979}. Examples of
such disclinations have been shown in living systems such as protein
coats of viruses \cite{iorio_virus_2007} and insect corneal
nanostructures\cite{lee_mesostructure_2016}, where insertion of
pentagons into an array of packed hexagons results in Gaussian
curvature and breaks the sixfold rotational symmetry. Similarly, the
Stone-Wales defect, characterised by two pentagons and two heptagons\cite{stone_theoretical_1986}, is a well-known defect in graphene,
fullerene, and carbon nanotubes\cite{ma_stone-wales_2009,heggie_stonewales_2016}.

Paper wasps of {\it{Polistes}} genus build nests hanging from one or
more stalks. The nest itself is made up of open hexagonal cells.
Variations in nesting architectures within paper wasps have been
attributed to selection pressures such as predation from ants or
flying insects, insulation from heat, and economic use of building
material\cite{eberhard_social_1969, jeanne_adaptiveness_1975,
  jeanne_latitudinal_1979,seeley_regulation_1981,
  wenzel_evolution_1991}.  The diversity of nest forms in paper wasps
and their growth have been elegantly explained by a simple set of
rules by Karsai and P\'enzes\cite{karsai_nest_1998}.  The model
suggests that multiple forms could be built by changing the weightage
assigned to different parts of the nest. However, by construction,
these models describe regular hexagonal structures only. The
likelihood of these models explaining changes to regular hexagonal
tiling or within nest changes to order is small. In this paper,
analysing the nesting architecture of paper wasps ({\it{Polistes
    wattii}}), we ask the following questions:
\begin{itemize}

\item[(a)] What is the nature of ordering (i.e., short-range vs
  long-range) and spatial arrangement of cells within and between the
  nests? In the current context, order describes the regularity in the
  tiling of the cells within the nests. Short-range order (SRO)
  represents the arrangements of nearest neighbours in the nest, while
  long-range order (LRO) represents the regularity over a longer
  distance. More quantitatively, a nest is said to have long-range
  order if there is a nonzero probability of finding the corners of
  two cells on the same regular lattice, even for large separations
  between the cells.

\item[(b)]{{What are the topological defects found in theses nests?}} 
\end{itemize}

\section*{Methods}
The study was conducted within the campus of Ashoka University,
Sonepat, India (28.9457$^{\circ}$ N, 77.1021$^{\circ}$ E) located
at an altitude of 224 m. The maximum temperature in summers is
45-47$^{\circ}$C, and the lowest temperature in winters is around
4-6$^{\circ}$C. Wasps of the genus {\it{Polistes}} have a
widespread distribution and are a well-studied example for the
evolution of sociality and dominance hierarchies in insects. It is a
primitive and speciose genus with more than 200 species\cite{carpenter_phylogeny_1996}.  {\it{P.wattii}} is known to be
distributed across central and South Asia\cite{ceccolini_new_2019}.
They are inactive during the winter, and nest building begins once
they emerge from hibernation. The construction of nests often lasts
till late summer. In our study, all the analysed {\it{P.wattii}} nests
were from anthropogenic habitats. Nesting height was typically
within the range of 5-15 m.  Images of nests were taken in the summer
of 2020 and 2021. Adult nests that were fully constructed and
completed the season were considered for analysis (typically with
at least 50 cells, Figure S4). These nests were photographed with a
reference scale. Images were further analysed by subtracting the
background and binarisation using ImageJ\cite{schneider_nih_2012}.
The positions of all the individual vertices (corners shared by
at least three cells) were marked on the images, and their x, y
coordinates were exported for subsequent analysis. \tr{Here, we have taken the structure to be planar (see Results and Discussion for details).} To avoid the
effect of boundary which includes incomplete cells, a few layers of
vertices along the edges were excluded. Apart from these, a few cells
with cylindrical projections where identifying the underlying vertices
was difficult were excluded from the final analyses. Overall, 25 nests
were used to characterise the degree of ordering. The presence of
topological defects was identified by inspecting the number of
vertices for the cells within 35 different nests. We characterised the defects
by identifying their basic features (number of vertices and
neighbouring cells) and calculating the internal angles of the first
neighbouring hexagonal layer from coordinates. For a region without
defects, we would expect these angles to be close to 120$^{\circ}$.

To quantify ordering and analyse the spatial distribution of cells, we
used pair correlation function $g(r)$ (henceforth PCF) based on Ripley's K function derivative method in the spatstat package with default smoothing parameter\cite{baddeley_spatial_2015}. The approach relies on the probability
of finding a pair of vertices separated by a distance $r$. This
results in an average representation of the local spatial
neighbourhood at a distance $r$ from any given vertex. PCF of a
regular lattice exhibits sharp peaks at the lattice spacing distances (resolved using a lower smoothing parameter than the default value). Such sharp peaks indicate perfect ordering. On the other hand, an
amorphous structure will show some reduced degree of order; in
particular, it would have $g(r)$ approaching 1 for large values of
$r$. The limiting value of 1 indicates no correlation between the
positions of vertices when they are far apart. In a PCF, the peak
heights are generally related to the number of neighbours. The first
two peaks were discernible in nests of different sizes, allowing us to
focus on the location and width of these peaks for each of the nests.
The following function consisting of two Gaussians with amplitudes
$a_1, a_2$ has been used to fit the first two peaks of the
$g(r)$-vs-$r$ curve for each nest,
\begin{eqnarray}
g_{\rm fit}(r)= a_1\, \exp\left({-\frac{(r-r_1)^2}{2s_1^2}}\right) 
     +a_2\, \exp\left({-\frac{(r-r_2)^2}{2s_2^2}}\right ),\label{eq:1}
\end{eqnarray}
where $r_1, s_1$ represent the location and the width of the first
peak, and $r_2, s_2$ of the second peak.  We also calculated the
coefficient of variation (CV) of location and width for the first two
peaks to quantify the extent of variation in order between nests.

\begin{figure}[htbp]
    \centering
    \includegraphics[width=0.6\linewidth]{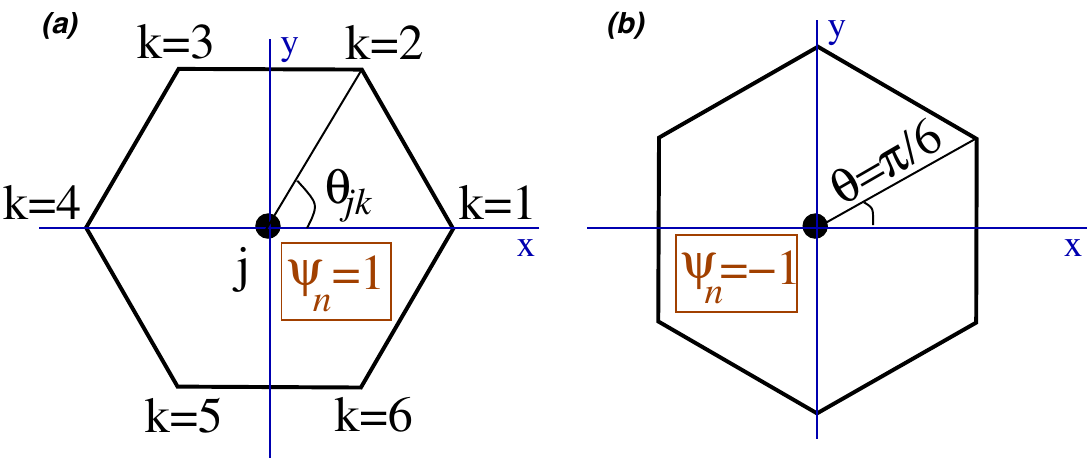}
    \caption{Cell-orientation order parameter (Eq. \ref{eq:2}). The
      angle $\theta_{jk}$ is measured from an arbitrarily chosen
      x-axis through the centre as shown in (a).  For a regular
      hexagon with inner angles at $2\pi/3$, the orientation in (a)
      has $\psi_n = 1$ while in (b) $\psi_n=-1$.  In general, $\psi_n$
      is orientation-dependent with $|\psi_n|$ less than or equal to
      1.  For regular hexagons, individual $\psi_n$ lies on the unit
      circle. The six-fold symmetry of a regular hexagon is preserved
      by $\psi_n$ as it returns to its value under a $\pi/3$ rotation
      of the hexagon around its centre.}
    \label{Figure 1}
\end{figure}

We measured the individual cell areas and compared the variation
between and within the analysed nests. We also calculated the nearest
neighbour distances or bond lengths, i.e., distances between vertices
(henceforth wall lengths). Based on the the fitted values of Eq.
\ref{eq:1}, the wall length would be in the range $r_1\pm 2s_1$.
Information of translational ordering is given by the PCF, and to
obtain the orientational order, we defined cell-orientational order in
analogy with bond-orientational order in liquids. Cell-orientational
order parameter is a measure of the geometrical arrangement of the
vertices of a given cell around its cell centre\cite{steinhardt_bond-orientational_1983,schilling_monte_2005}.
Cell-orientational order parameter $\psi_n$ is given by a complex
number defined at each cell centre as,
\begin{equation}
    \psi_{n}(\mathbf{X}_{j})=\frac{1}{N_{j}}\sum^{N_{j}}_{k=1} e^{in\theta_{jk}},
    \label{eq:2}
\end{equation}
where $N_j$ is the number of vertices of cell $j$ centered at
${\mathbf{X}_j}$, $\theta_{jk}$ represents the angle between the line
joining vertex $k$ to the centre and the chosen x-axis, and $n$ is an
index for orientational ordering (Figure \ref{Figure 1}).
For hexagonal ordering, we choose $n=6$. Note that $\psi_n$ neither
depends on the vertex numbering scheme nor the size of the hexagon.
As a complex number, $\psi_n=|\psi_n| e^{i \theta}$ lies on or within
the unit circle in the complex plane, and therefore,
$\psi_{n}(\mathbf{X}_{j})$ can be represented graphically by a vector
at an angle $\theta$ with the x-axis at the cell centre
$\mathbf{X}_{j}$.  The average $\langle \psi_n\rangle$ of
$\psi_{n}(\mathbf{X}_{j})$ over a large number of hexagons lies in the
range $0\leq |\langle \psi_n\rangle|\leq 1$ with the two extreme
values representing a random orientational arrangement 
($\langle\psi_6\rangle=0$) and a perfect hexagonal ordering
($\langle\psi_6\rangle=1$).   We calculated the cell-orientational order
parameter for a simulated set of random central angles to study the
decay of $|\langle\psi_6\rangle|$ from the perfect value of 1 (note that $\langle\psi_6\rangle=0$ corresponds to  independent hexagons which do not form a regular lattice; details in Supplementary Information). Analyses were performed by using
Fortran (https://gcc.gnu.org) and R software\cite{r_core_team_r_2020}.

\section*{Results and Discussion}
\subsection*{Nature of ordering and spatial arrangement of cells in the nests}
\begin{figure}[htbp]
    \centering
    \includegraphics[width=0.8\linewidth]{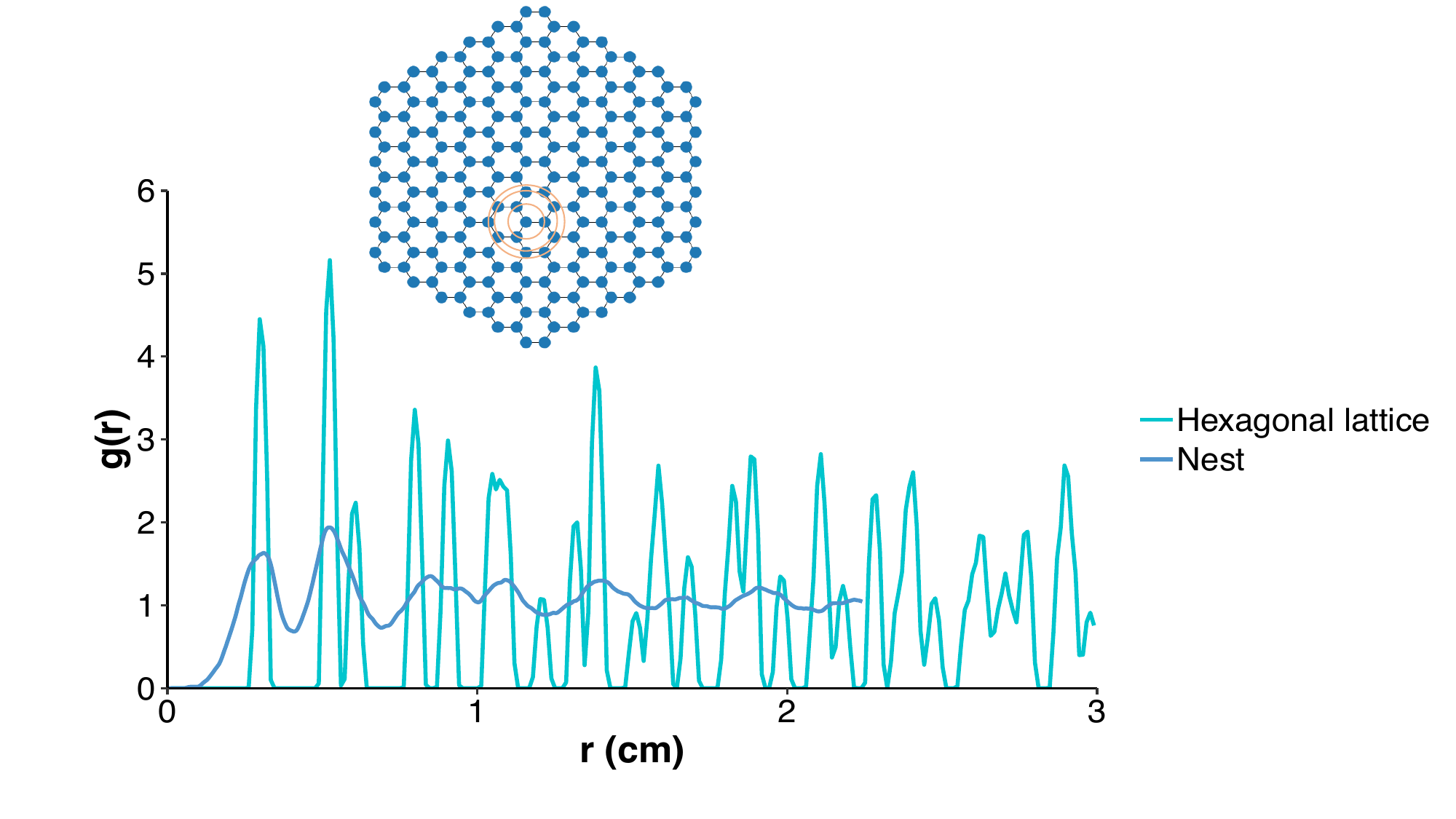}
    \caption{Pair correlation function as a function of $r$, i.e., the
      distance between vertices { {in cm. The two curves depict a
      representative nest and a regular hexagonal lattice. The regular hexagonal lattice has been constructed with a wall length equal to the average wall length of the representative nest. A schematic on the top shows a regular
      hexagonal lattice and the vertices in the first, second, and
      third shells around any one vertex. Each vertex is connected to three other vertices in the first shell.}}}
    \label{Figure 2}
\end{figure}
The number of cells in the analysed nests ranged between 50 and 740. For each nest, we characterised the nature of ordering by calculating pair correlation function for the vertices. For a regular hexagonal lattice, sharp peaks (in principle, $\delta$-function) occur at distances corresponding to radii of different shells indicating the presence of both short and long-range order\cite{bishop_pair_1984}.  As shown in Figure \ref{Figure 2}, for any vertex on the hexagonal lattice, the first, second, and third neighbours lie on circular shells and the first 3 PCF peaks reflect this regularity. We found that there is a substantial overlap between the first and second peaks of a regular hexagonal lattice and those of a typical nest. As $r$ increases, this overlap disappears. This pattern of the PCF illustrates the SRO in the structure of nests. 
The presence of a peak in a PCF confirms that any given vertex would have neighbouring vertices at distances provided by the peak position. This is an example of SRO. Moreover, the decaying envelope of the PCF to $g(r)=1$ is a signature of the absence of LRO. The proximity of the third peak to the second peak is an inherent characteristic of a hexagonal lattice, suggesting that the two shells are closeby. However, the second shell and third shell are not resolved in the case of nests, \tr{due to broad wall-length distribution. The absence of sharp peaks and the missing proximity of second and third shells are signatures of deviation from regularity.} Indeed, the broad peaks in the nest PCF curves indicate heterogeneity of cell sizes that impairs ordering over long distances. Therefore, we conclude that similar to amorphous structures, nests exhibit SRO but no LRO. 

\begin{figure}[H]
    \centering
    \includegraphics[width=0.8\linewidth]{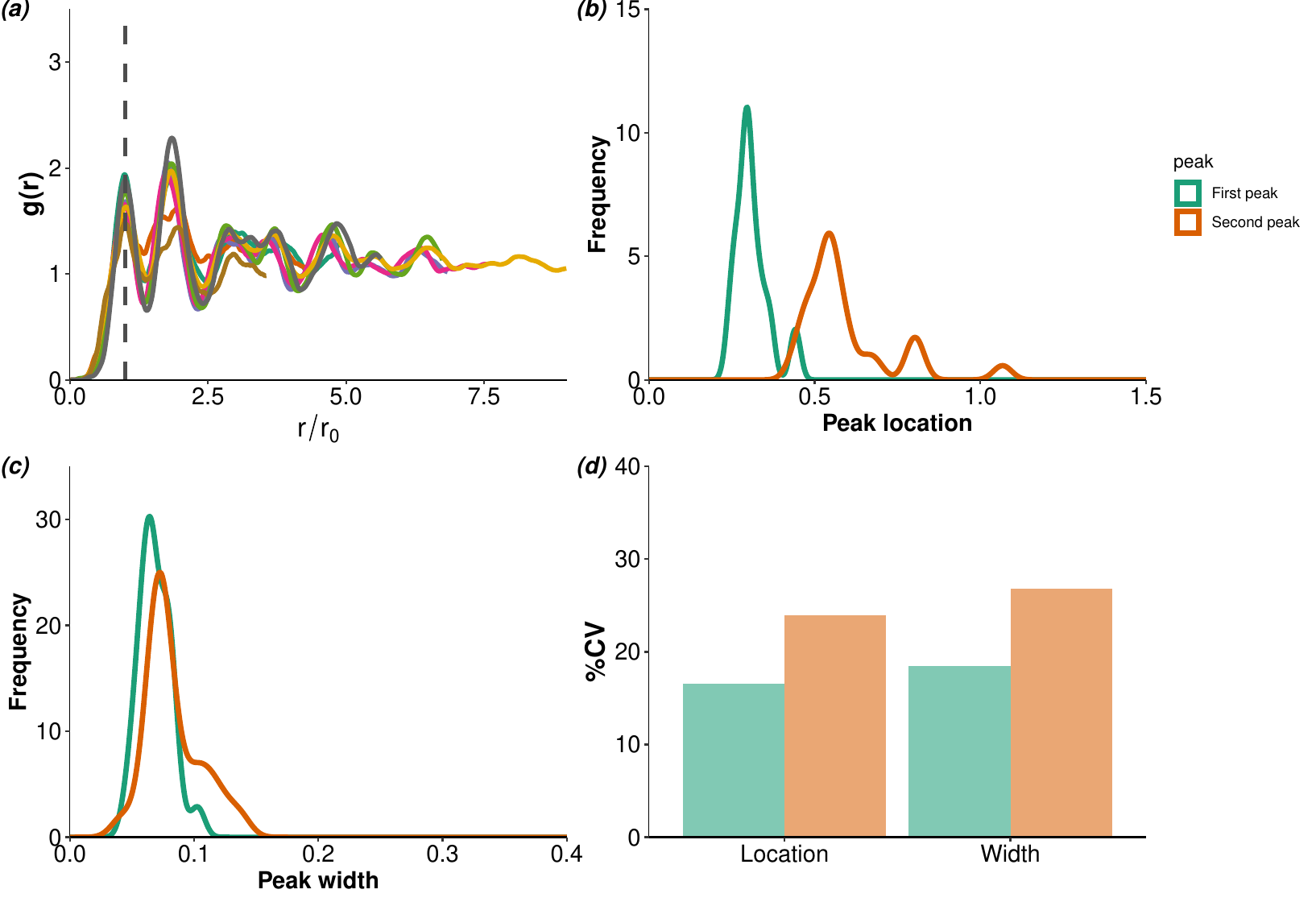}
    \caption{(a) Pair correlation function of vertices from 10
      randomly chosen nests, showing a significant overlap around the
      first and second peaks; beyond the third peak, the positioning
      of vertices approaches random distribution (approaching 1). The
      x-axis is scaled ($r/r_0$) such that the first peak is at the
      same x-value for all nests. {{Each coloured line represents one nest.}} Distribution of (b) locations and
      (c) widths obtained by Gaussian approximation of first and
      second peaks of the pair correlation function of 25 nests. (d)
      Relative difference in coefficient of variation (CV) of
      locations and widths of first and second peaks. Each CV value
      depicts variation within the nest.}
   \label{Figure 3}
\end{figure}

The PCF curves of nests shown in Figure \ref{Figure 3}(a) have been normalized such that the first peak is at $r=1$ (non-normalized individual nests' PCFs are shown in Supplementary Information, Figure S1). When variations in translational order of nests were analysed, the nest-to-nest variation was found to be restricted to larger $r$ values. This is
further corroborated by the low variance values (coefficient of variation, CV) of location and width of the first peak compared to the second peak (Figure \ref{Figure 3}(b), (c), (d)).

\begin{figure}[t]
    \centering
    \includegraphics [width=0.6\linewidth]{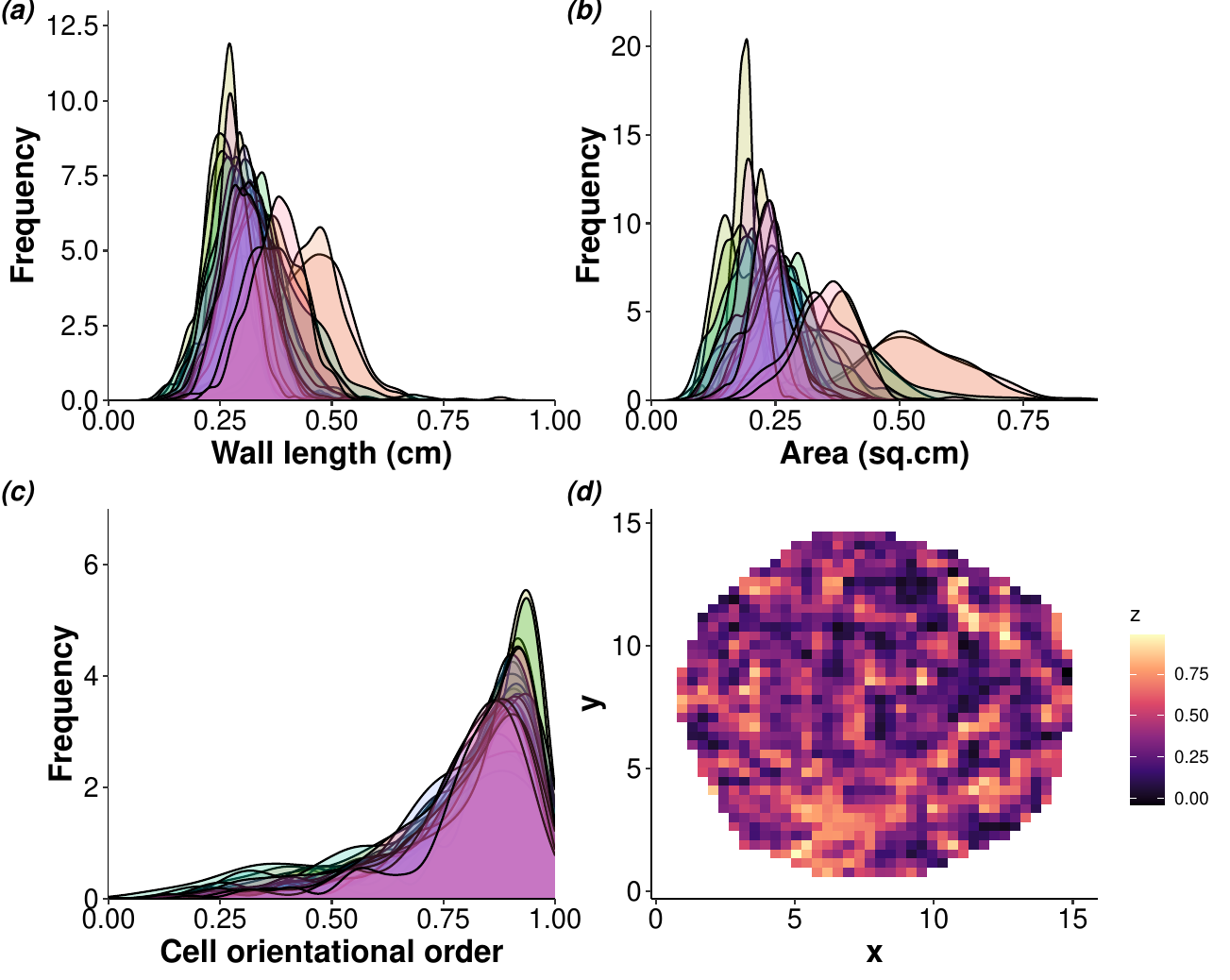}
    \caption{Frequency distribution of (a) cell wall lengths, (b) cell
      areas, and (c) cell orientational order parameter { {(dimensionless), depicting
      variation between and within the nests. (d) Heat map showing $z={\rm Im} \; \psi_6$,
      the imaginary part of $\psi_6$  of cell orientational order parameter of a representative nest ($x$,$y$ in cm indicate centre of each cell).}}}
    \label{Figure 4}
\end{figure}

The structural features were analysed by characterizing the distribution of wall lengths, cell areas, and cell orientational order (Figure \ref{Figure 4}). Figure \ref{Figure 4}(a) depicts the extent of variation in wall lengths between the different nests. The lengths
varied within the range of 0.1 and 0.9 cm, with a median value of 0.3 cm. The cell areas were in the range of 0.07 to 0.85 sq.cm (median= 0.25 sq.cm, Figure \ref{Figure 4}(b)). \tr{The distribution of cell wall-lengths also fits well with the
values obtained from the Gaussian fit of the first peak of the PCF. Furthermore, we analysed if the degree of deviation was greater within the nest or between the nests. Percent CV values of cell areas ranged between 13 and 28 while the wall-lengths varied from 44 to 58\%. The wall-length variation was greater within the nests than between the nests. Cell wall length distributions were analyzed for randomly chosen cells from central and peripheral regions of wasp nests to test the assumption of planarity of the nest. For a curved surface, one expects the wall lengths of peripheral cells to be different from those in the central region. These distributions were found to overlap (Figure S5). Therefore, we conclude that the curvature-induced systematic distortion (if any) is indistinguishable from the wall-length distribution of the cells}. 

Our analysis suggests that the orientation of the cells is fairly preserved and aligns well with the nearest neighbours (Figure \ref{Figure 4}(c), (d)). The orientational order parameter varied between the nests in the range of 0.1 and 1, with a median value of 0.85. Significant overlap in cell orientational order parameter distributions of the nests suggests that the presence of orientational order is true across the nests (Figure \ref{Figure 4}(c)). Percent CV values varied between 16 and 31, indicating that { {variation within nests in cell orientational order was smaller than the variation in wall lengths.}} Compared to the case of a single cell with random internal lengths (see Supplementary Information), the probability distribution of $\langle\psi_6\rangle$ in Figure \ref{Figure 4}(c) is wider. We \tr{attribute this to the} cooperative effect of building a compact nest and yet becoming amorphous-like in the large scale limit. A well-studied example of such cooperative-effect induced broadening is the probability distribution of magnetism as one approaches the Curie temperature of a magnet\cite{fleury_phase_1981}. Taken together, the pair correlation function and distribution of cell orientational order parameters suggest the presence of short-range hexagonal order and an overall orientational order even though there is no long-range order akin to amorphous materials.

\subsection*{Topological defects}
Hexagonal structures are prone to defects that emerge in a variety of ways. In the analysed nests, we show the presence of topological defects in the form of non-hexagonal cells. Akin to defects in most materials, these were found to occur at low frequency (20\% of the nests). In few nests of {\it{P. annularis}}\cite{wenzel_endogenous_1989}, { {non-hexagonal cells were reported to induce the required curvature of nests.}} However, the presence of octagons and higher-order defects has not been reported.
 
\begin{figure}[htbp]
    \centering
    \includegraphics[width=0.5\linewidth]{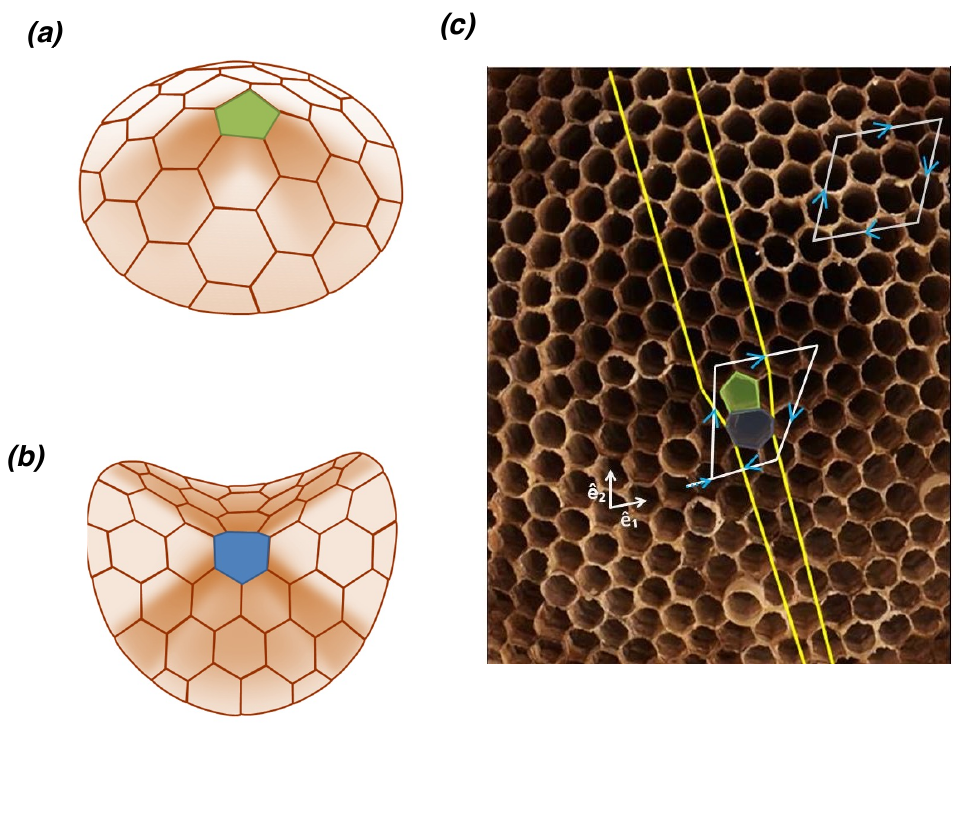}
    \caption{Illustration of emergent structure that deviates from
      planarity with the insertion of (a) a pentagon or (b) a
      \tra{heptagon (shading as a guide to the eye).} (c) A disclination pair is indicated on a wasp nest. A
      view of parallel lines encompassing an additional hexagonal
      layer as they pass through this pair, thereby leaving a scar on
      the lattice (dislocation line). A Burgers vector obtained from a
      loop enclosing the dislocation is a topological invariant. A
      loop not enclosing a defect is shown on the top right.
      $\widehat{\bf e}_1$ and $\widehat{\bf e}_2$ are the basis
      vectors.}
    \label{Figure 5}
\end{figure}

\begin{figure}[htbp]
    \centering
    \includegraphics[width=0.8\linewidth]{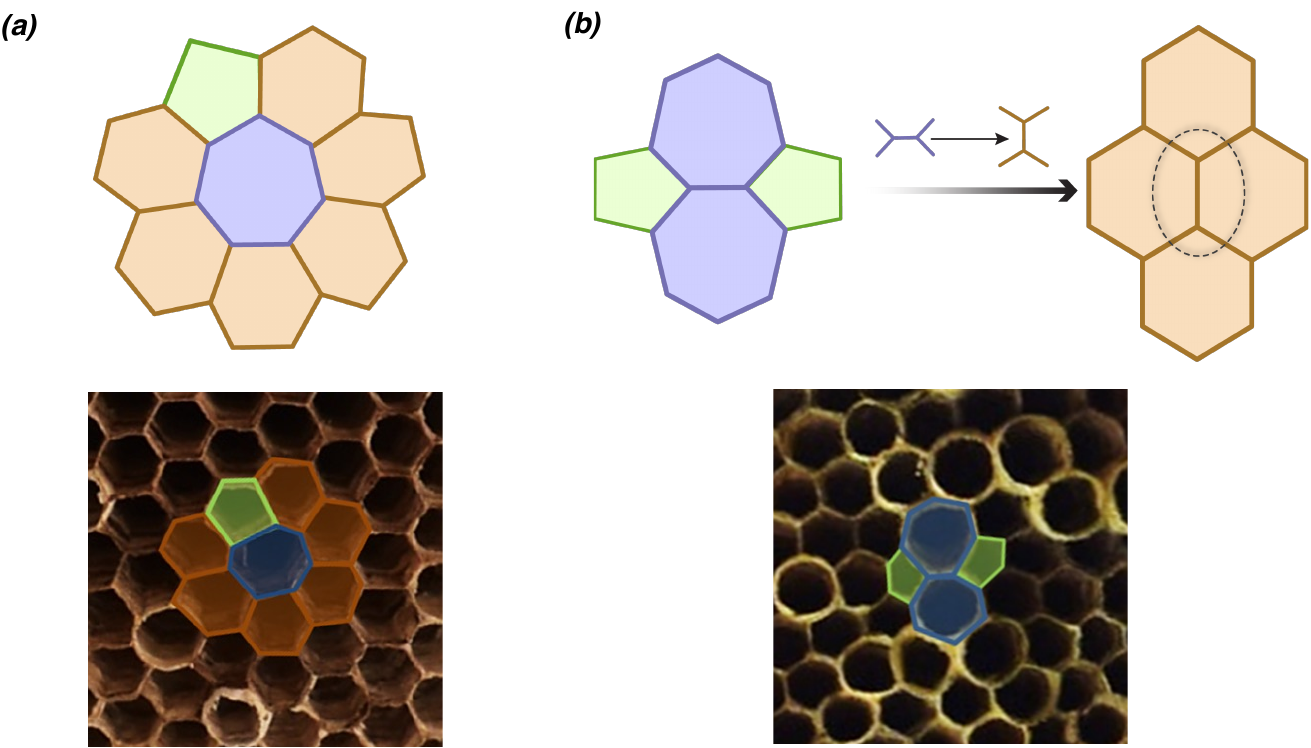}
    \caption{Schematic representation of (a) a pentagon and a heptagon
      amidst hexagons (dipole), and (b) Stone-Wales defect
      (pentagon-heptagon quadrupole) \tra{or a dislocation dipole.} The arrow points towards the representation of hexagons when these defects are fixed. The possible modifications to the wall are indicated above the arrow. Bottom panels show their outline %
      in the wasp nests.}
    \label{Figure 6}
\end{figure}

\begin{figure}[htbp]
    \centering
    \includegraphics[width=0.9\linewidth]{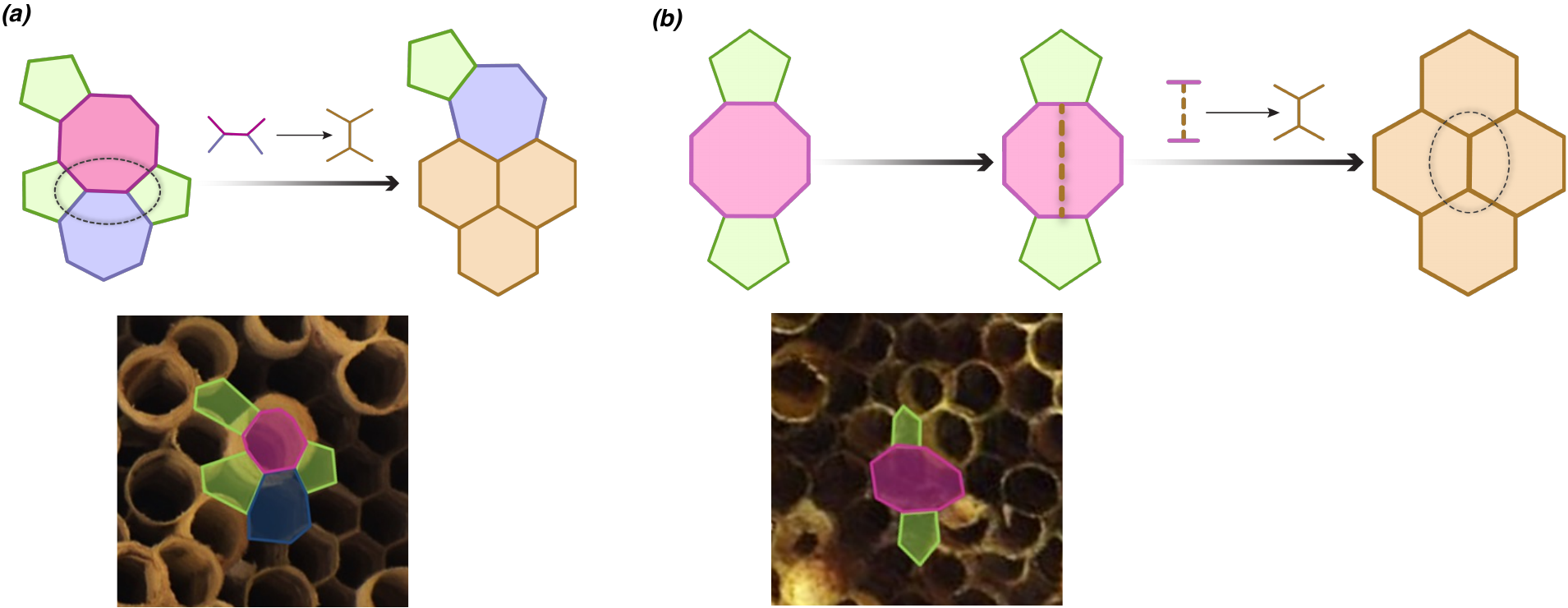}
    \caption{Schematic representation of (a) an octagon surrounded by
      a heptagon and three pentagons, and (b) an octagon between two
      pentagons. The arrows point towards the representations of hexagons when these defects are fixed. The possible modifications to the wall are indicated above the arrow. Bottom panels show their outline in the wasp nests.}
    \label{Figure 7}
\end{figure}

 \tr{Defects such as one pentagon or one heptagon amidst hexagons disrupt the planarity of a honeycomb structure as shown in} Figure \ref{Figure 5}(a,b)\cite{zsoldos_effect_2010}. \tr{If structural disorders such as topological defects are left uncorrected, they affect the overall planarity of the nest}. The missing link of a pentagon in a hexagonal net necessarily requires the removal of a wedge of hexagons, producing a cone-like structure as shown in Figure \ref{Figure 5}(a). Similarly, an opposite curvature can be produced by a single heptagon (Figure \ref{Figure 5}(b)) with an extra link that requires the
insertion of a wedge. As these curvatures cannot be ironed out, these
are topological defects called disclinations that tend to disrupt the
orientational order of the lattice. A pair of the two opposite types
do not cancel each other but instead carry with them a line defect,
which is called a dislocation line. This dislocation line consists of
an extra array of hexagonal cells (Figure \ref{Figure 5}(c)).
Furthermore, \tr{a dislocation is characterised} by drawing a loop in a region without defects and comparing it with a similar loop that encloses it. The steps required to close the loop are
characteristic of the defect and independent of the size and shape of
the loop. This topological invariant called Burgers vector is shown in Figure \ref{Figure 5}(c). 
Figures \ref{Figure 6} and \ref{Figure 7} show the distinct types of such defects found within the analyzed nests. Stone-Wales defect comprises of two pentagons and two heptagons (Figure \ref{Figure 6}(b)), \tr{which typically arises by a simple 90$^{\circ}$ rotation of the bond (wall) between two hexagons. As a pentagon-heptagon pair is a dipole, Stone-Wales defect becomes equivalent to a disclination quadrupole (two oppositely oriented dipoles). These nomenclatures are analogous to those of electric charges. Though a local fault in construction can produce a Stone-Wales defect}(Figure \ref{Figure 6}(b)), \tr{it is still topological in nature as the distortion propagates to the boundary}. Such
topological defects are known to occur commonly in graphite and graphene\cite{ophus_large-scale_2015}. For cells with defects that occur in pairs or as groups, we measured the distribution angles of the neighbouring vertices for upto one shell of nearest neighbours.
For a regular hexagonal lattice, these angles would be
120$^{\circ}$, and for the nest without defects, we found that a set of randomly
chosen cells subtended angles that ranged between 90$^{\circ}$ and
140$^{\circ}$.  However, hexagonal cells adjoining the defect pairs
such as pentagon-heptagon showed angles that range from 99$^{\circ}$
to 151$^{\circ}$, which differed from both the regular hexagonal lattice as well as hexagonal cells in the nest. Figure \ref{Figure 7}(a)
depicts a defect where an octagon is embedded within three pentagons,
four hexagons, and a heptagon. This \tr{is obtained by a
90$^{\circ}$ rotation} of the wall between three hexagons and a
heptagon. Another defect that was observed was an octagon with two
pentagons on either side (Figure \ref{Figure 7}). This \tr{complex changes to a
four-hexagon configuration by a wall addition followed by a split
(change in the wall angles of the central wall)}. Such extended defects are equivalent to pairs of dislocations, which \tr{do not have any associated Burgers vector. These arise in scenarios where wasp(s) have constructed the outline of cells, and the internal walls were added subsequently}. All the
above-described defects were found to maintain the three-point
vertices (vertex connected by three walls).


A dislocation for a two-dimensional system consists of a pentagon-heptagon pair (i.e., a pair of opposite disclinations). \tra{The presence of such a defect necessarily entails an extra row of hexagons that vitiates the translational order at large distances but not  the
bond-orientational order. Consequently, a proliferation of dislocations can produce a state of  bond-orientational order without any translational order (a distinct phase of matter well known as the hexatic phase). A liquid is formed when the dislocations break into disclinations (independent pentagons and heptagons){\cite{hexatic}}, which destroy the residual bond-orientational order. For instance, when these defects are induced by thermal energy, one sees a two-step melting process of a two dimensional crystal. In colloidal crystals, dislocations or disclination dominated phases can be produced by external agencies like laser.} In short, topological defects significantly influence a system's behavior.   In this spirit,  the disposition of the two-dimensional nests is ultimately determined by the nature of the individual cells and the topological defects, \tra{which are generated by the construction rules adopted by the wasps}.
 
While pentagon-heptagon pairs were found to occur in nests, the configuration where hexagons separate a pentagon and a
heptagon was never observed in our dataset. Interestingly, some of
the described defects and their properties are utilized by honeybees
to attain curvature as well as to merge cells of different sizes\cite{hepburn_construction_1991,smith_imperfect_2021}. Cells of
different sizes are routinely observed in honeycombs as larvae of
drones and workers are of different sizes. Retaining order with cells
of different sizes that are being built independently by different
individuals poses a challenge, and the honeybees counter this by local
sensing and introducing non-hexagonal cells where {{required}}\cite{nazzi_hexagonal_2016, smith_imperfect_2021,tarnai_buckling_1989}. \tr{The introduction of a pentagon or a heptagon
adjacent to each other is only possible by a regular set of local
inspections carried out by wasps to identify possible deviations from
hexagonality. There are two alternatives to explain the emergence of these defects that require further investigations.  These alternatives are as follows: a) As the angle of a regular heptagon is 128.5$^{\circ}$ is closer to that of a regular hexagon than a pentagon whose
angle is 105$^{\circ}$, heptagons have arisen first, which are then fixed immediately by adding a
pentagon, and  b) the higher number of pentagons
overall suggests that moulding cells with lesser material could
have erroneously resulted in pentagons that were subsequently fixed by
inserting a heptagon.} It is likely that experience and learning play a
crucial role in repair processes. Experiments in {\it{P. fuscatus}}
have shown the presence of an elaborate building programme that is
beyond a set of linear steps, including repairing cell walls when
damaged, adding pulp, and strengthening the stalk when required, which
is done by multiple inspections\cite{downing_regulation_1990}. This
suggests that deviations from 6-sidedness are also possibly assessed
via such inspections. However, it is unknown if these inspections
involve identifying the number of vertices, measuring the internal
angles or by Burgers vector.

\section*{Conclusion}
Our results suggest that the nests of {\it Polistes wattii} show orientational order without any translational long-range order in the placement of the hexagonal cells on a planar surface.  However, the ordering is well preserved in the immediate neighborhood of the cells (short-range order). Furthermore, topological defects were identified in the form of dislocations and disclinations (non-hexagonal cells) in the nests. We showed that these defects were organized such that the planarity of the nests was maintained. We used the approach of pair correlation function and topological methods to analyse these structural features of wasp nests.

Two key features that were never violated in the nests were a)
convexity of individual cells and b) planar global structure. \tr{We
postulate that wasps conduct systematic localized inspections, detect changes in local symmetry or order, and make changes to the adjoining cells to restore order.} Going further, we intend to test the role of cell size heterogeneity on the
emergence of topological defects and the relationship between the decay length of pair correlation function with the size of the nests. Future studies could
place the nests in captivity and characterise the behaviours that
precede and succeed the construction of cells around the defects. The
introduction of defects experimentally in such captive nests would
help understand the response of individuals and the group. While our
study establishes the presence and repair of the defects, it is
unclear if all the wasps equally participate in this process or
certain keystone individuals are efficient at inspection and repair
processes. These additional studies would therefore allow us to
understand the behavioral mechanisms encompassing the repair of topological defects.

\medskip

\noindent\textbf{Data availability}\\ 
 The datasets used in the current study will be available from the corresponding author on request.

\noindent\textbf{Acknowledgements}\\
We thank Sreya Dey for taking up this idea in its pilot phase and
initiating the work. Thanks to Smruti Ranjan and Abdelsalam Gena for
help with data analysis. We thank Upasana Sengupta, Diana Michael, and
Nageshwar Kumar for help with collecting nests.  

\medskip
\noindent\textbf{Author contributions}\\
S.K. conceived and designed the study. A.G. and SK collected the data. S.K. and S.M.B. performed and interpreted the analysis and drafted the
manuscript. All authors gave final approval for publication.

\medskip
\noindent\textbf{Funding}\\
S.K. acknowledges intramural funding from Ashoka University. A.G. thanks Science and Engineering Research Board, India (CRG/2019/003297) for fellowship. S.M.B. thanks JC Bose Fellowship (SR/S2/JCB-71/2009) from Science and Engineering Research Board, India. 

\medskip
\noindent\textbf{Competing interests}\\
The authors declare no competing interests.



\section*{Supplementary material}

# Ordering and topological defects in social wasps' nests
# Supplementary Information

Shivani Krishna[1,*], Apoorva Gopinath[1] and Somendra M. Bhattacharjee[2]

[1]Department of Biology, Ashoka University, Sonepat 131029, India
[2]Department of Physics, Ashoka University, Sonepat 131029, India
*Corresponding author: shivani.krishna@ashoka.edu.in

## S1. Plots of pair correlation functions

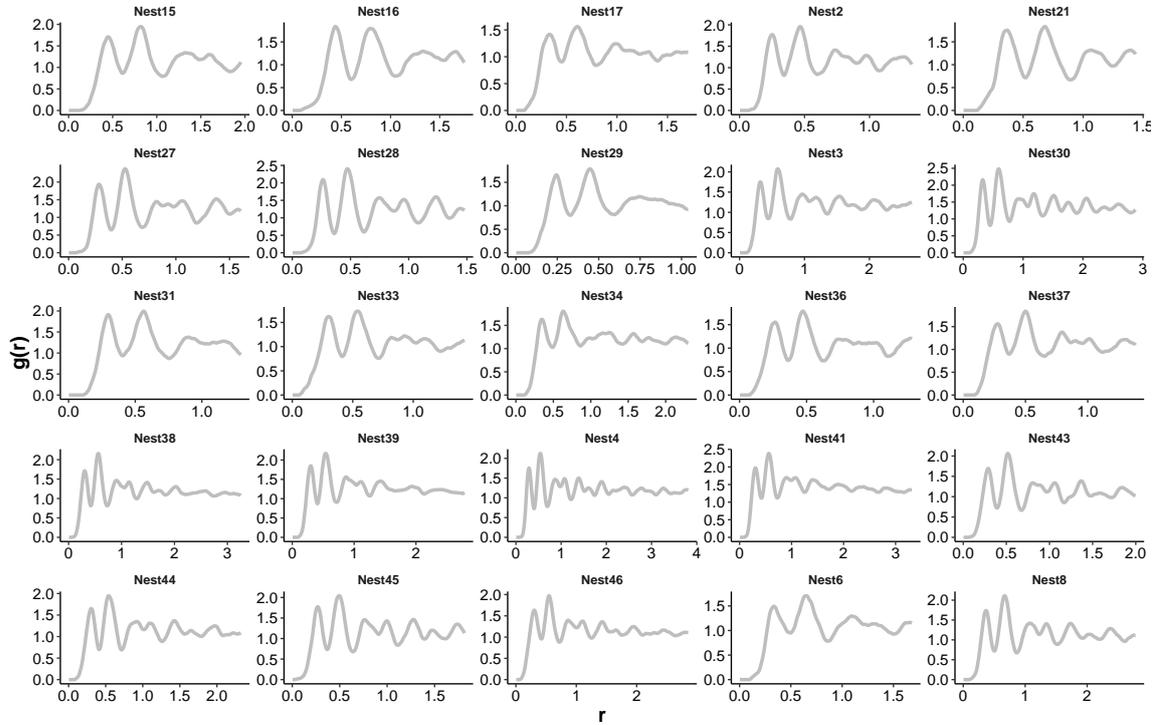

Figure S1: Pair correlation functions of vertex positions for each of the analysed nests. With increasing $r$ (the separation between two vertices, in cm), the curve tends to approach 1 across all the nests.

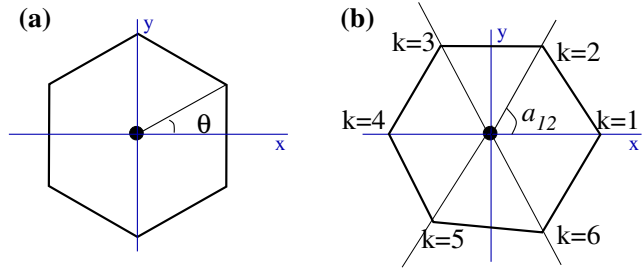

Figure S2: Orientations of a hexagon. (a) A regular hexagon with an orientation angle $\theta$ with respect to the arbitrarily chosen x-axis. For a given $\theta$, $\langle\psi_6\rangle = e^{i6\theta}$, with $0 \leq \theta \leq \pi/3$. (b) Irregular hexagon constructed by randomly choosing the angles $a_{kl}$.

## S2. Random orientation—A hexagon with different types of randomness

**Random Orientations**

Consider a regular hexagon as shown in Figure S2(a), with $\theta$ representing the overall orientation with respect to the x-axis. Suppose the orientation angle $\theta$ is random, uniformly distributed in the range $[0, \pi/3)$. The range is dictated by the symmetry that the hexagon returns to an indistinguishable position under a rotation by $\pi/3$. The normalized probability distribution of $\theta$ is,

$$P(\theta) = \frac{3}{\pi}, \quad \text{with} \quad \int_0^{\pi/3} P(\theta)\, d\theta = 1. \tag{S1}$$

For a given $\theta$, the successive angles of a regular hexagon are given by

$$\theta_k = (k-1)\frac{\pi}{3} + \theta, \quad k = 1,...6,$$

so that the cell-orientational order-parameter is

$$\psi_6(\theta) = \frac{1}{6}\sum_k e^{i\theta_k} = e^{i6\theta}.$$

Therefore, the average is given by,

$$\langle\psi_6\rangle = \frac{3}{\pi}\int_0^{\pi/3} e^{i6\theta} d\theta = 0. \tag{S2}$$

This is because for every orientation $\theta < \pi/6$ with $\psi_6 = e^{i6\theta}$, there is an equally probable orientation $(\pi/6)+\theta$ with $\psi_6 = -e^{i6\theta}$, and they cancel each other. In other words, for a disordered situation where a hexagon can take any orientation at random, the average order parameter is zero as mentioned in the text.

## S3. Random angles

Refer to Figure S2(b). Suppose each of the angles $a_{kk+1}, (k = 1,...,5)$, subtended by the sides of an irregular hexagon at the cell center is chosen randomly (independent and identically distributed)

in the range $(\pi/3) \pm A$ with a probability distribution $P(a) = \frac{1}{2A}$. The value of $A$ is kept small to ensure that $a_{61} > 0$. Let us choose the x-axis such that the vertex numbered 1 is on the axis as in Figure S2(b). The polar angle for site $k$ is,

$$\theta_k = \sum_{j=1}^{k-1} a_{jj+1}, \quad k = 2,...,6, \tag{S3}$$

and

$$\langle e^{-i6a} \rangle = \frac{1}{2A} \int_{-A}^{A} e^{i6a} \, da \approx 1 - 6A^2, \quad \text{(for small } A\text{)} \tag{S4}$$

Using Eq. S3, we evaluate the average cell-orientational order parameter in terms of $X = \langle e^{-i6a} \rangle$ as,

$$\langle \psi_6 \rangle = \frac{1}{6} \left(1 + X + X^2 + X^3 + X^4 + X^5\right), \tag{S5}$$

which for small $A$ can be written as (from Eq. S4)

$$\langle \psi_6 \rangle \approx \frac{1}{6} \sum_{j=0}^{5} (1 - 6A^2)^j \approx e^{-15A^2}. \tag{S6}$$

Figure S3 shows a comparison of the approximate formula for small $A$ (Eq. (S6)) with numerical simulations. The agreement is good for small $A$.

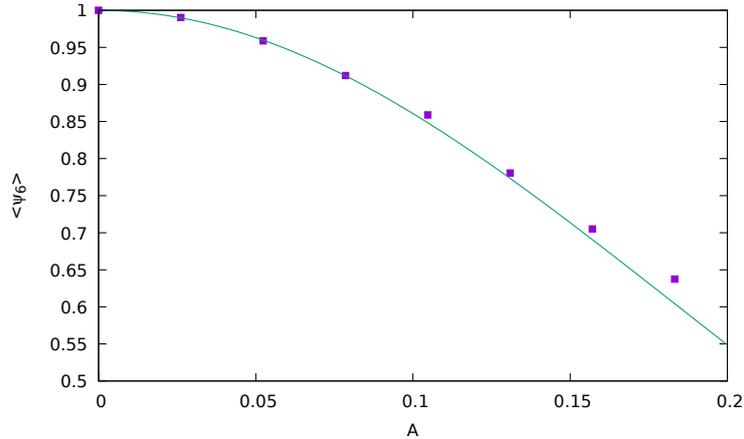

Figure S3: Randomness in irregular hexagon: $\langle \psi_6 \rangle$ vs $A$, where the central angles are uniformly distributed in $(\pi/3 - A, \pi/3 + A)$. Results from numerical simulation using uniform random numbers are shown by the filled squares, while the solid line represents the exponential formula of Eq. (S6).

# S4. A sample nest

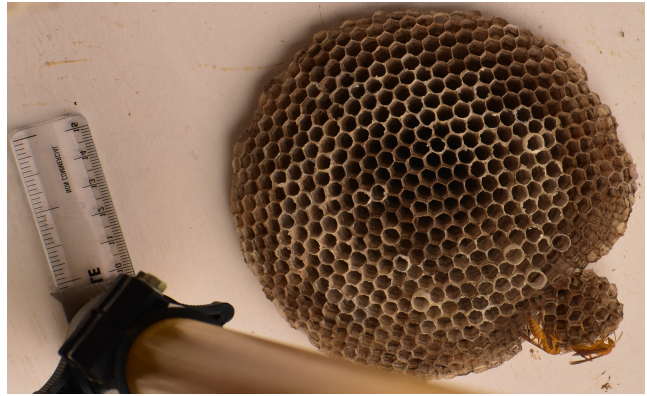

Figure S4: Nest of *Polistes wattii*.

# S5. On planarity of nests

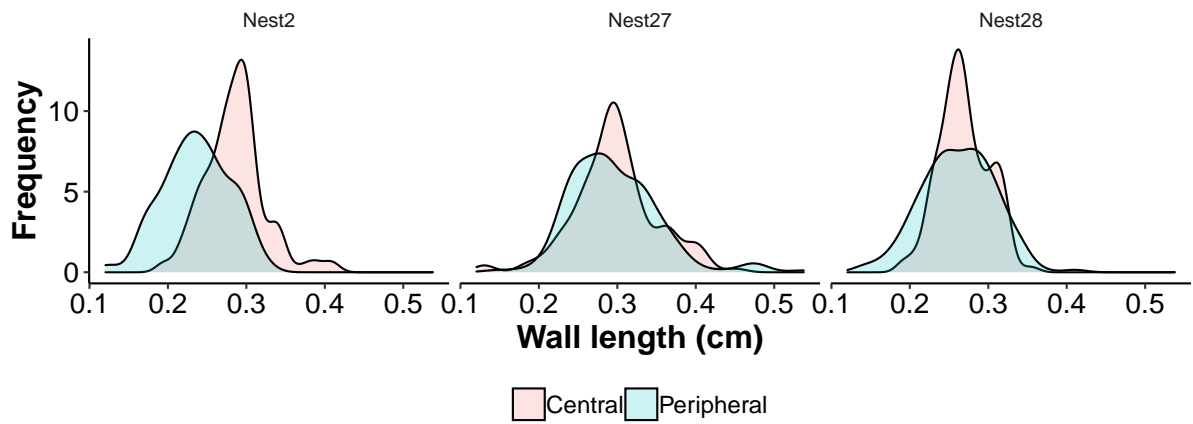

Figure S5: Comparison of wall length distribution between randomly chosen cells that lie in the central and the peripheral regions of three different nests



\end{document}